\documentclass[conference]{IEEEtran}
\IEEEoverridecommandlockouts

\usepackage{listings}
\usepackage{caption}
\usepackage{xcolor}
\usepackage{hyperref}   

\newcommand{\quantemrepo}{\url{https://github.com/quinn-langfitt/QuantEM}}

\usepackage{url}
\usepackage{cite}
\usepackage{amsmath,amssymb,amsfonts}
\usepackage{algorithmic}
\usepackage{graphicx}
\usepackage{textcomp}
\usepackage{xcolor}
\def\BibTeX{{\rm B\kern-.05em{\sc i\kern-.025em b}\kern-.08em
    T\kern-.1667em\lower.7ex\hbox{E}\kern-.125emX}}
\begin{document}
\title{QuantEM: The quantum error management compiler}

\author{\IEEEauthorblockN{Ji Liu$^1$, Quinn Langfitt$^2$,  Mingyoung Jessica Jeng$^2$, Alvin Gonzales$^1$, Noble Agyeman-Bobie$^3$, Kaiya Jones$^4$  \\ Siddharth Vijaymurugan$^1$, Daniel Dilley$^1$, Zain H. Saleem$^1$, Nikos Hardavellas$^{2,5}$, Kaitlin N. Smith$^{2}$}
\IEEEauthorblockA{
\textit{$^1$Mathematics and Computer Science Division, Argonne National Laboratory, Lemont, IL, USA} \\ \textit{$^2$Department of Computer Science, Northwestern University, Evanston, IL, USA} \\
\textit{
$^3$Department of Computer Science, Grambling State University, Grambling, LA, USA} \\ 
\textit{
$^4$Department of Computer Science, Tuskegee University, Tuskegee, AL, USA} \\ 
\textit{$^5$Department of Electrical and Computer Engineering, Northwestern University, Evanston, IL, USA}
}
}

\maketitle

\begin{abstract}
As quantum computing advances toward fault-tolerant architectures, quantum error detection (QED) has emerged as a practical and scalable intermediate strategy in the transition from error mitigation to full error correction. By identifying and discarding faulty runs rather than correcting them, QED enables improved reliability with significantly lower overhead. Applying QED to arbitrary quantum circuits remains challenging, however, because of the need for manual insertion of detection subcircuits, ancilla allocation, and hardware-specific mapping and scheduling.

We present QuantEM, a modular and extensible compiler designed to automate the integration of QED codes into arbitrary quantum programs. Our compiler consists of three key modules: (1) program analysis and transformation module to examine quantum programs in a QED-aware context and introduce checks and ancilla qubits, (2) error detection code integration module to map augmented circuits onto specific hardware backends, and (3) postprocessing and resource management for measurement results postprocessing and resource-efficient estimation techniques. The compiler accepts a high-level quantum circuit, a chosen error detection code, and a target hardware topology and then produces an optimized and executable circuit. It can also automatically select an appropriate detection code for the user based on circuit structure and resource estimates. QuantEM currently supports Pauli check sandwiching and Iceberg codes and is designed to support future QED schemes and hardware targets. By automating the complex QED compilation flow, this work reduces developer burden, enables fast code exploration, and ensures consistent and correct application of detection logic across architectures.
\end{abstract}


\section{Introduction}
Quantum computing holds great promise for addressing practical problems in areas such as chemistry simulation~\cite{robledo2025chemistry_beyond_exact_diag}, machine learning~\cite{Liu_2021ArigorousAndRobustQuantSpeedupInSupervisedML}, combinatorial optimization~\cite{farhi2014quantum-QAOA}, and cryptography~\cite{Bennett_2014BB84}. Recent experimental demonstrations~\cite{Kim_2023EvidenceOfUtilForQCBeforeFaultTol} have illustrated the potential of quantum hardware in tackling problems that are challenging to simulate on classical supercomputers. Many of these demonstrations rely on quantum error mitigation techniques such as probabilistic error cancellation~\cite{Temme_2017ErrorMitigForShortDepthQuantCircs} and zero-noise extrapolation~\cite{Li_2017ZNEeffVarQuantSimIncIAEM, Temme_2017ErrorMitigForShortDepthQuantCircs}. As the field transitions from error mitigation toward fully developed error correction, quantum error detection (QED) emerges as a critical intermediate approach. Instead of correcting errors, error detection identifies faulty runs and discards them, providing a useful compromise during this transitional stage of hardware capability.


As quantum computers scale up, error mitigation approaches become less effective due to their exponential sampling overhead. \cite{Takagi_2022, Quek_2024}. For instance, ZNE relies on circuit scaling to suppress errors, but this quickly leads to circuit sizes that are too large to sample efficiently, making accurate estimation impractical at higher scaling factors. In contrast, QED shifts some of this overhead to additional qubits and gates, thereby reducing the reliance on excessive sampling.

While QED offers significant potential for mitigating noise, however, applying it to arbitrary circuits remains challenging. It requires expertise in code selection and syndrome extraction along with an understanding of the hardware-specific constraints. Moreover, the addition of error detection subcircuits and ancilla qubits introduces overhead and sources of additional noise. As a result, circuits augmented with QED must be carefully designed to balance the trade-off between maximizing the benefits of error detection and minimizing the noise and complexity introduced by the added circuit components. Without tooling, a developer must manually modify circuits to allocate ancilla qubits, insert checks, and route measurements. This process is time-consuming, error-prone, and difficult to scale. Fortunately, carefully designed software and compilers can help alleviate the challenges associated with injecting QED into quantum circuits.

Quantum circuit compilers help translate quantum circuits expressed in high-level abstractions into specifications that are quantum processor ready through context-aware optimization. Prior art in quantum compilation has tackled challenges such as placement and routing in communication-constrained hardware~\cite{li2019tackling,smith2019quantum}, navigating heterogeneous system noise~\cite{murali2019noise,tannu2019not}, leveraging pulse level control~\cite{gokhale2020optimized, campbell2023superstaq}, and mapping to modular quantum systems~\cite{baker2020time,jeng2025modular,bandic2025profiling}, among others. However, existing Noisy Intermediate-Scale Quantum (NISQ) era compilers are not designed to account for the structure and constraints of error detection subcircuits. When such subcircuits are treated as part of the original program, standard compiler optimizations can interfere with the intended error propagation paths (for instance, routing passes of the compiler can naively insert SWAP operations that pass over qubits monitoring noise), potentially resulting in incorrect or ineffective circuits. These challenges highlight the need for a dedicated quantum compiler that explicitly distinguishes between the original computational circuit and the error detection subcircuits. Such a compiler should be able to correctly and efficiently integrate error detection into arbitrary circuits while preserving their intended functionality and mapping them to the constraints of the target hardware.

We propose the \underline{Quant}um \underline{E}rror \underline{M}anagement compiler, QuantEM\footnote{\quantemrepo}. QuantEM automates the task of placing QED into arbitrary quantum circuits. The compiler accepts a high-level quantum program and the topology of a target hardware, and can either incorporate a user-specified detection code or automatically select an appropriate strategy based on circuit structure and resource estimates. It analyzes the algorithm to locate critical error regions, selects appropriate points to insert checks, and produces a low-level circuit with the required ancilla qubits and measurement operations. By maintaining a library of detection codes and backend mappings, the compiler allows users to swap codes and hardware targets with minimal effort. Our current compiler supports the compilation of Pauli check sandwiching (PCS)~\cite{Debroy_2020ExtendedFlagGadgetsForLowOverCircVer, gonzales2023quantum_PCS, liu_2022classicalsimulatorsquantumerrorMitigViaCircCut, Li_2024Qutracer, langfitt_2024paulicheckextrapolationquantum, langfitt_2024dynamicresourceallocationquantumWErrDetect} and Iceberg~\cite{self2024protecting_iceberg} error detection codes. This work reduces developer burden, enables rapid code exploration, and ensures consistent application of detection primitives.




\section{Background}

Error detection codes add extra qubits and measurements to flag runtime errors.  The simplest example is the three-qubit repetition code, which detects bit-flip errors by measuring a pair of parity checks.  More general stabilizer codes detect arbitrary Pauli errors by measuring generators of the code stabilizer group.  Each code has specific requirements on connectivity, gate set, and measurement timing.

\begin{figure}[t]
     \centering
         \includegraphics[width=0.99\linewidth,trim={0cm .5cm .5cm .5cm},clip]{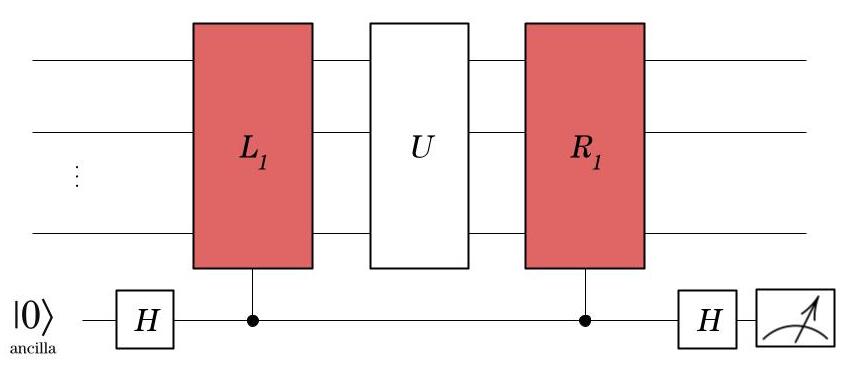}
        \caption{General PCS circuit layout. The red unitaries represent the Pauli checks that sandwich the main payload circuit that acts on the target qubits. Measurement of the ancillas provide detection of errors that occurred in the payload circuit.}
        \label{fig:pcs-circuit}
        
\end{figure}

Pauli check sandwiching (PCS) is an error detection scheme that provides protection to the targeted qubits. As seen in Figure~\ref{fig:pcs-circuit}, PCS surrounds a payload circuit acting on the targeted qubits, $U$, with controlled Pauli operator checks. Errors on $U$ can be detected on an ancilla through phase kickback.
It is important that the relationship 

\begin{equation}\label{eq:pauli-check-facts}
    R_{1}UL_{1} = U
\end{equation}

\noindent holds so the introduced checks do not disturb the original payload circuit semantics. As a note, the operations contained in $L_1$ and $R_1$ do not have to comprise the same Pauli gates as long as Eqn.~\ref{eq:pauli-check-facts} is satisfied. PCS can be considered a distance 1 code since it generally cannot detect arbitrary single-qubit errors. However, PCS forms a distance 2 code, which is locally unitarily equivalent to a CSS code, when appropriate checks are used \cite{Gonzales_2025DetectErrsInAQuantNetWithPauliCh}. Martiel and Javadi--Abhari ~\cite{martiel2025low_PCS} recently formalized the problem for identifying valid Pauli checks by converting it to a linear code decoding problem and applying heuristic methods for its solution. They also introduced a low-overhead check placement scheme that first maps the circuit to the physical hardware and then places ancilla checks adjacent to the corresponding physical qubits.

The Iceberg code is a distance 2 code, $[[k+2, k, 2]]$ for even $k$, that scales efficiently with the number of logical qubits. The Iceberg code  requires only two additional qubits for the encoded state. The code implements fault-tolerant initial state preparation and syndrome measurement circuits capable of detecting any single-qubit error. Because its structure involves two physical qubits connected to all others, it is ideally suited to the trapped-ion computers with all-to-all connectivity.

Compilers for quantum circuits typically focus on mapping logical gates to hardware constraints and optimizing depth.  Few systems integrate error detection as a first-class transformation.  Prior work has demonstrated software stacks for inserting checks for PCS~\cite{cirstoiu2023volumetric_QERMIT_PCS} and Iceberg~\cite{jin2025iceberg_compiler, qiskit-iceberg-transpiler}; they have shown that these two QED techniques are promising for compiling applications. However,  no general-purpose compiler  supports multiple codes and backends.

\section{Compiler Framework}

\begin{figure}
    \centering
    \includegraphics[width=0.9\linewidth]{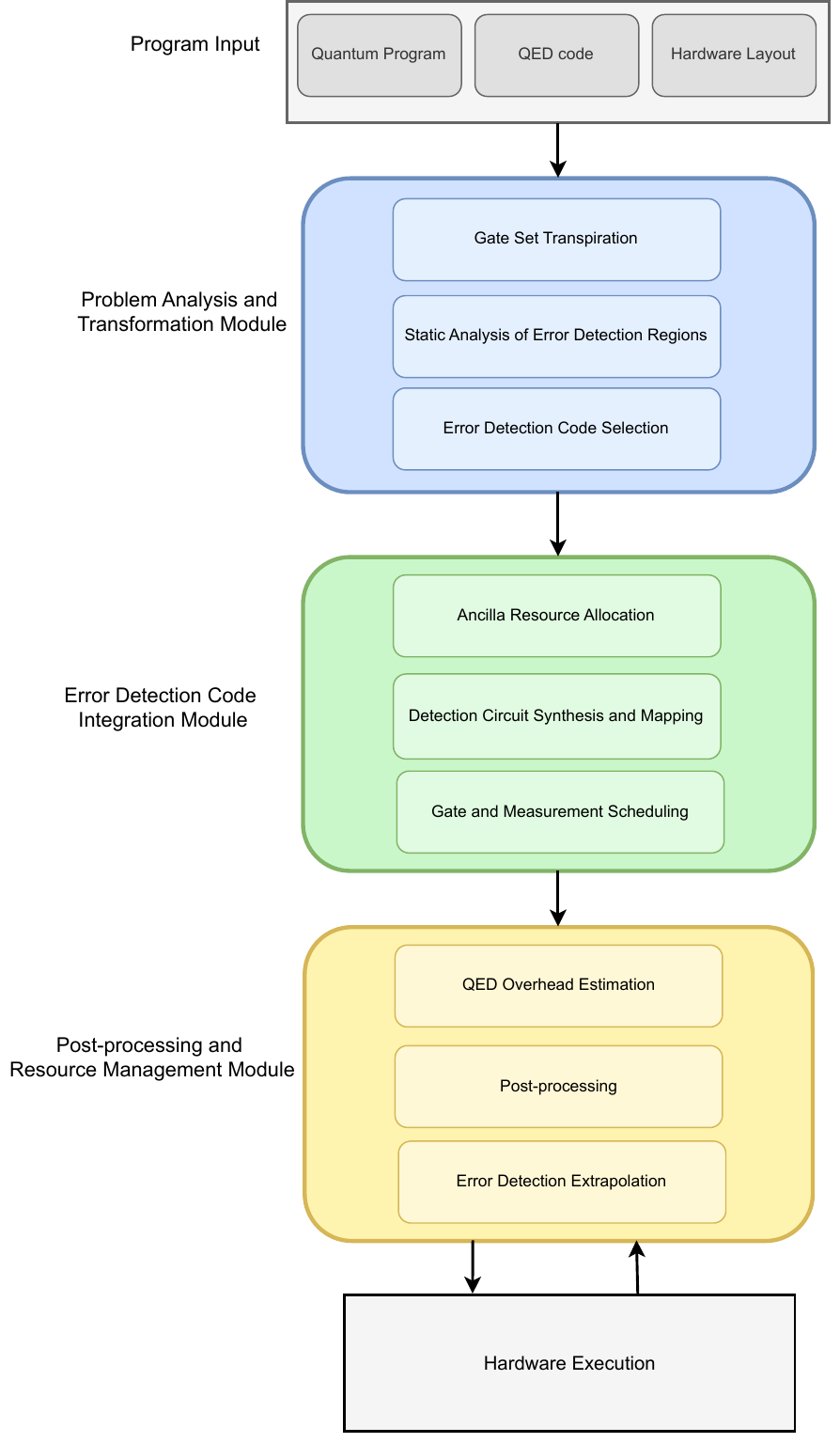}
    \caption{Modular compiler pipeline for quantum error detection}
    \label{fig:QED-compiler}
\end{figure}

We propose a modular and extensible compiler QuantEM~\cite{github_repo} for compiling quantum circuits with QED codes. The workflow of our modular compiler is shown in Figure~\ref{fig:QED-compiler}.  Each module consists of multiple passes, enabling flexible composition and customization of the compilation pipeline. We list several key passes in the figure. 
Our proposed compiler contains three key modules. 

\textbf{Program Analysis and Transformation Module}: The module tackles the program analysis at the logical circuit level, transpiles the circuit to a target gate set with QED protection, and then selects appropriate error detection codes based on the circuit structure. Since the analysis is performed at the logical circuit level, it is not constrained by hardware limitations, allowing flexibility in exploring the design space for inserting error detection checks. Moreover, this enables co-design opportunities between error detection code insertion and qubit mapping algorithms. The primary passes in this module include:

\begin{itemize}
    \item \textbf{Gate Set Transpilation}: Converts the input quantum circuit into a target gate set that aligns with the hardware or detection strategy requirements.

    \item \textbf{Static Analysis of Error Detection Region}: Performs static analysis of the circuit to identify blocks of operations that are suitable candidates for error detection. These may include high-fidelity subcircuits, idle regions, or segments with long coherence durations.

    \item \textbf{Error Detection Code Selection}: Chooses an appropriate detection code for the quantum circuit. For example, if a circuit contains  large Clifford subcircuits, this pass selects the PCS code over alternatives such as the Iceberg code, which may be more suited to circuits with many two-qubit Pauli rotations.
\end{itemize}

The program analysis and transformation module identifies the most suitable QED code for protecting the circuit. It also annotates qubits and gates with metadata to guide subsequent insertion of detection logic.

\textbf{Error Detection Code Integration Module}: The module allocates ancilla qubits for QED, synthesizes the error detection subcircuit, maps the circuit to target hardware, and schedules the gates and measurements. 
The integration process ensures that the inserted detection logic is compatible with hardware constraints and does not interfere with the original computation. The main compilation passes for this module are:

\begin{itemize}
    \item \textbf{Ancilla Resource Allocation}: Inserts the ancilla qubits and performs initial logical-to-physical qubit mapping.

    \item \textbf{Detection Circuit Synthesis and Mapping}: Synthesizes the error detection subcircuit based on the selected code, and performs qubit mapping and routing to satisfy hardware connectivity constraints. This pass prioritizes the mapping of the error detection subcircuits and minimizes the extra SWAP operations inserted.

    \item \textbf{Gate and Measurement Scheduling}: Schedules the gates to maximize the parallelism and minimze the idling time. It also schedules the measurements on the ancilla qubits to enable qubit reuse.
\end{itemize}

This error detection code integration module highlights the need for a dedicated compiler, since standard compilers do not differentiate between gates from the original circuit and those introduced for error detection. 

\textbf{Postprocessing and Resource Management Module}: The module handles QED-specific overhead estimation and postprocessing of measurement results and performs resource-efficient estimation techniques such as Pauli check extrapolation~\cite{langfitt_2024paulicheckextrapolationquantum}. Its functionality is organized into the following passes:

\begin{itemize}
    \item \textbf{QED Overhead Estimation}: Estimates the overhead associated with postselection, considering the number of qubits, hardware noise profile, and circuit structure. This pass may also incorporate the device execution results to perform more accurate estimation. For instance, the pass incorporates the ancilla-free Pauli checks~\cite{langfitt_2024dynamicresourceallocationquantumWErrDetect} for estimating the noise profile.

    \item \textbf{Postprocessing}: Postprocesses the circuit’s measurement results based on ancilla qubit outcomes, using detection syndromes defined by the chosen QED code.

    \item \textbf{Error Detection Extrapolation}:  Based on the number of ancilla qubits and observed detection patterns,  can optionally apply check extrapolation techniques~\cite{langfitt_2024paulicheckextrapolationquantum} to estimate the final expected outcome.
\end{itemize}

This module automates the postselection process and incorporates extrapolation strategies to manage the trade-off between detection accuracy and resource overhead, enabling scalable use of QED.

\section{Case Study: QED Insertion and Compiler Workflow}
\subsection{PCS example}
In this section, we demonstrate an end-to-end compilation workflow using a 4-qubit circuit with the QuanTEM compiler. The code showcases the three core modules of the compiler. 
First, in the \textbf{Problem Analysis and Transformation} module, the input circuit is processed by the \texttt{convert\_to\_PCS\_circ\_largest\_clifford()} function, which inserts Pauli checks into the largest Clifford subcircuit. 

Second, in the \textbf{Error Detection Code Integration} module, the circuit is mapped to a target backend using the \texttt{get\_VF2\_layouts()} function, which finds a suitable layout based on the backend’s topology. The mapped circuit is then transpiled with Qiskit to match the basis gate set of the backend. 

Third, after simulating execution on the fake backend, the \textbf{Postprocessing} module applies \texttt{postselect\_counts()} to filter measurement results based on the ancilla qubit outcomes.
\begin{minipage}{\linewidth}
\begin{lstlisting}[caption={PCS circuit generation}, captionpos=t]
from quantem.utils import (
    convert_to_PCS_circ_largest_clifford,
    get_VF2_layouts,
)
from quantem.pauli_checks import postselect_counts
from qiskit_ibm_runtime.fake_provider import FakeWashingtonV2

num_qubits = 4
num_checks = 2

# Create Fake backend based on device properties
fake_backend = FakeWashingtonV2()

# Problem Analysis and Transformation Module
sign_list, pcs_circ = convert_to_PCS_circ_largest_clifford(circ, num_qubits, num_checks)

# Error Detection Code Integration Module
# Map to backend using VF2 algorithm
pcs_VF2_mapping_ranges, pcs_small_qc = get_VF2_layouts(pcs_circ, fake_backend)

pcs_circ_transpiled = transpile(
    pcs_circ,
    basis_gates=basis_gates,
    initial_layout=pcs_mapping_range,
    optimization_level=1,
)

# Circuit Execution
pcs_circ_counts = noisy_sampler.run(
    [pcs_circ_transpiled], shots=10000
).result().data.meas.get_counts()

# Post-processing Module
postselected_counts = postselect_counts(pcs_circ_counts, num_ancillas=2)
\end{lstlisting}
\end{minipage}

\subsection{Iceberg code example}
In this section, we demonstrate an end-to-end compilation workflow using a 6-qubit QAOA circuit with the QuanTEM compiler. The code showcases the core modules of the compiler.
First, in the \textbf{Problem Analysis and Transformation} module, the input circuit is processed by the \texttt{build\_iceberg\_circuit()} function, which inserts state-preparation circuits and syndrome measurement circuits.  

Second, we create a custom noise model with a fully connected backend and the depolarizing noise for single- and two-qubit gates. 

Third, after simulating execution on the fake backend, the \textbf{Postprocessing} module applies \texttt{postselect\_counts\_iceberg()} to filter measurement results based on the ancilla qubit outcomes.

\begin{minipage}{\linewidth}
\begin{lstlisting}[caption={Iceberg circuit generation}]
from quantem.iceberg_code import build_iceberg_circuit
from qiskit_aer import AerSimulator

num_qubits = 6
num_checks = 2

# Problem Analysis and Transformation Module
iceberg_circ, reg_bundle = build_iceberg_circuit(
    qaoa_circ,
    optimize_level=3,
    attach_readout=True,
    total_syndrome_cycles = num_checks
)

# Create a custom noise model
noise_model = NoiseModel()
# Add depolarizing error to all single and two qubit gates.
error = depolarizing_error(0.002, 2)
noise_model.add_all_qubit_quantum_error(error, ["cx", "rzz", "rxx", "ryy"])
error = depolarizing_error(0.00003, 1)
noise_model.add_all_qubit_quantum_error(error, ["u1", "u2", 'u3'])

# Create a fakebackend with the noise model
fakebackend = AerSimulator(noise_model=noise_model)

#Circuit Execution
iceberg_circ_counts = noisy_sampler.run(
    [iceberg_circ], shots=10000
).result().data.meas.get_counts()

fakebackend.run(qed_qc, shots = 100000).result()

# Post-processing Module
postselected_counts = postselect_counts_iceberg(iceberg_circ_counts, num_ancillas=2)
\end{lstlisting}
\end{minipage}

\section{Conclusion}
In this paper, we propose QuantEM, a modular and extensible compiler that automates the integration of QED codes into quantum programs. Our compiler accepts a high-level quantum circuit and a target hardware backend and then produces an optimized and executable circuit. QuantEM reduces developer burden and enables user-friendly application of QED techniques to quantum algorithms.  As new hardware and techniques emerge, we plan to update QuantEM as needed.

\section*{Code Availability}
QuantEM is available at \quantemrepo

\section*{Acknowledgments}
This material is based upon work supported by the U.S. Department of Energy, Office of Science, National Quantum Information Science Research Centers.
JL, AG, and ZHS acknowledge support by
the Q-NEXT Center. AG, DD, and ZHS also acknowledge support by the U.S. Department of Energy (DOE) under Contract No. DE-AC02-06CH11357, through the Office of Science, Office of
Advanced Scientific Computing Research (ASCR) Exploratory Research for Extreme-Scale Science and Accelerated Research in Quantum Computing. NH was partially supported by NSF CCF-2119069. This work used resources awarded by the IBM Quantum Credits program. We acknowledge the use of IBM Quantum services. The views expressed are those of the authors, and do not reflect the official policy or position of IBM or the IBM Quantum team.

\vspace{1em}
\framebox{\parbox{0.93\linewidth}{
The submitted manuscript has been created by UChicago Argonne, LLC, Operator of Argonne National Laboratory (“Argonne”). Argonne, a U.S. Department of Energy Office of Science laboratory, is operated under Contract No. DE-AC02-06CH11357. The U.S. Government retains for itself, and others acting on its behalf, a paid-up nonexclusive, irrevocable worldwide license in said article to reproduce, prepare derivative works, distribute copies to the public, and perform publicly and display publicly, by or on behalf of the Government.  The Department of Energy will provide public access to these results of federally sponsored research in accordance with the DOE Public Access Plan. http://energy.gov/downloads/doe-public-access-plan.}}

\bibliographystyle{plain}  
\bibliography{refs}

\end{document}